\documentclass[checkin,pre,showpacs]{revtex4}
\begin{document}        %
\draft
\title{Comments on "Pathway from Condensation via Fragmentation to Fermionization of Cold Bosonic Systems"} 
\author{A. Kwang-Hua CHU \thanks{Present Address : P.O. Box 39, Tou-Di-Ban, Xihong Road,
Urumqi 830000, PR China.  }}  
\affiliation{P.O. Box 30-15, Shanghai 200030, PR China}
\begin{abstract}
We make remarks on the paper by   Alon and  Cederbaum [{\it Phys. Rev. Lett.}
{\bf 95} 140402 (2005)]. The information about the
Pauli blocking for Fermi-Dirac and Bose-Einstein particle statistics were presented.

\end{abstract}
\maketitle
\doublerulesep=6mm    %
\baselineskip=6mm
\bibliographystyle{plain}               
Alon and  Cederbaum just presented a pathway from condensation to fermionization  where the evolution of the density profile of a cold bosonic system with increasing scattering length was followed [1]. With increasing scattering length the density
profile acquires more and more oscillations, until their number eventually equals to N, the number of bosons. Once the number of oscillations equals the number of bosons
and the system becomes fermionized.
They showed that the ground state and density profile of a bosonic system strongly
depend on the shape of the trap potential.
This allows one to design systems with an intriguing
ground state, e.g., for which one part is condensed and the other one is fermionized.
\newline
However, the present author,
based on the experiences in the quantum kinetic theory
(especially the Uehling-Uhlenbeck approach) [2], would like to
make some remarks about their claim : the pathway from condensation to
fermionization depends on the shape of the potential. When more particles are
trapped in the double-well trap employed there, the physical picture
obtained above remains essentially the same? Alon and  Cederbaum
followed the pathway from weakly- to strongly-interacting bosons in real
space and obtain a wavefunction picture of the ground state of cold atoms in the trap.
This will allow them to monitor directly the fine changes in the spatial
density of interacting bosons as the inter-particle interacting ($\lambda_0$)
increases (say,
$\lambda=\lambda_0 (N-1)/(2\pi)=1$, the energy functional is minimal when all atoms
reside in one orbital, thus recovering Gross-Pitaevskii theory [1]; $N$ is the
number of bosons). In fact, Alon and  Cederbaum  chose as an illustrative 1D example
the potential profile $V (x)$ (cf. Fig. 1 in [1]) which was an asymmetric double-well potential with the left well being deeper and narrower than the right one.
With their numerical results, from $\lambda \approx 155$ (or $\gamma\equiv \lambda_0/2n \sim 9$) on, all bosons reside in different orthogonal orbitals-the bosonic system has become fully fermionized!
\newline
Considering the spin characteristic and the Pauli blocking
effect in quantum-mechanic sense [2-4], the above mentioned
results should better be attributed to the intrinsic
particle-statistics.  To remind the readers
with the  definition of Pauli blocking, we shall cite the relevant work below
for introducing the Uehling-Uhlenbeck equation [2-4], which extends
its applicability to particles which obey non-conventional
statistics. Consider particles (mass
m) which obey the Bose-Einstein or Fermi-Dirac statistics. These
particles interact elastically by means of binary encounters
(cross section $\sigma$). In the space homogeneous case the UUE
reads as follows ($\Lambda$=1 for bosons, $\Lambda$ = $-1$ for
fermions; $\Lambda$ is the parameter characterizes the Pauli blocking effect):
\begin{displaymath}
\frac{\partial f}{\partial t} = \frac{1}{h^3} \int d{\bf p}_1  d
{\bf \Sigma}' [f' f'_1 (1 + \Lambda f)(1 + \Lambda f_1)-f f_1(1 +
\Lambda f')(1 + \Lambda f'_1 )] \times g \sigma(g;\xi)) ;
\end{displaymath}
where $f = f({\bf p}; t)$ is the quantum distribution function
(dimensionless) and $f' = f({\bf p}'; t)$; $f_1=f({\bf p}_1; t)$;
$f'_1 =f({\bf p}'_1 ; t)$. We denote by ${\bf p}$ and ${\bf p}_1$
the momenta before collision, while ${\bf p}'$ and ${\bf p}'_1$
are the momenta after collision:
 $ {\bf p}' = ({\bf p}+{\bf p}_1+m g{\bf \Omega}')/2$;
 $ {\bf p}'_1 = ({\bf p}+{\bf p}_1-m g {\bf \Omega}')/2$ ;
where $g=|{\bf p}-{\bf p}_1|/m$ and ${\bf \Omega}'$ is a unit
vector. The cross section $\sigma$, in general, depends on both
$g$ and $\xi$, where $\xi ={\bf \Omega}\cdot {\bf \Omega}'$; ${\bf
\Omega} = 1/ ({\bf p}- {\bf p}_1)$. \newline The tuning of $\lambda$ in [1] is nothing
but the tuning of the rarefaction parameter in [4] together with
the Pauli-blocking parameter (say, $B$ in [4] or cf. Chu in [5]). There might be
resonances (or localizations), however, once $\lambda$ increases to
the order of magnitude of 1000 or more
[5]. For bosons, the parity of particle and antiparticle is the
same in contrast to fermions where the parity reverses between antiparticles.
The other issue is the effective mass for different systems characterized by, say, $\lambda$.
To conclude in brief, the Pauli blocking effect and the possible resonance
(or localization)
are crucial for those results in [1] which needs to be further checked.

\end{document}